\documentclass{ws-procs975x65}

\newcommand{\bea}{\begin{equation}\begin{array}{c}}
\newcommand{\eea}{\end{array}\end{equation}}
\newcommand{\ea}{\end{array}} 

\newcommand{\beq}{\begin{equation}}
\newcommand{\eeq}{\end{equation}}
\newcommand{\bad}{\begin{array}{ccc}}

\newcommand{\ba}{\begin{array}{c}}
\def\cropnote{\relax}

\crop[off]
\begin{document}

\vspace*{-12mm}
\begin{flushright}
SISSA 18/2015/FISI\\
IPMU15-0042~~~~~~~~~~\\
\end{flushright}
\vspace*{0.7cm}

\title{PREDICTIONS FOR THE DIRAC CP VIOLATION PHASE IN THE 
NEUTRINO MIXING MATRIX
}

\author{S. T. PETCOV
\footnote{Invited talk given at 
the International Conference on Massive Neutrinos, 
Nanyang Technological University, 
Singapore, 9-13 February, 2015; 
to be published in the Proceedings of the Conference.}
}

\address{SISSA/INFN, Trieste, Italy, and\\
Kavli IPMU (WPI), University of Tokyo, Tokyo, Japan}

\author{I. GIRARDI and A. V. TITOV}

\address{SISSA/INFN, Trieste, Italy}

\begin{abstract}
Using the fact that the neutrino mixing matrix
$U = U^\dagger_{e}U_{\nu}$, where $U_{e}$ and $U_{\nu}$
result from the diagonalisation of the charged lepton
and neutrino mass matrices, we analyse 
the predictions based on
the sum rules which
the Dirac phase $\delta$ present in $U$
satisfies when $U_{\nu}$ has a form dictated
by, or associated with, discrete flavour symmetries and
$U_e$ has a ``minimal'' form (in
terms of angles and phases it contains)
that can provide the requisite
corrections to $U_{\nu}$, so that the
reactor, atmospheric and solar neutrino mixing angles
$\theta_{13}$, $\theta_{23}$ and  $\theta_{12}$
have values compatible with the current data.
\end{abstract}


\vspace{-0.1cm}
\bodymatter
%
\section{Introduction}
\label{aba:sec1}
%

One of the major goals of the future
experimental studies in neutrino physics
is the searches for CP violation (CPV) effects
in neutrino oscillations
(see, e.g., 
Refs.~\refcite{Agashe:2014kda,Agarwalla:2013kaa,Adams:2013qkq,deGouvea:2013onf}).
It is part of a more general and ambitious program 
of research aiming to determine the status of the
CP symmetry in the lepton sector.

 In the case of the reference 3-neutrino mixing scheme,
CPV effects in the flavour
neutrino oscillations,
i.e., a difference between the probabilities of
$\nu_l \rightarrow \nu_{l'}$ and
$\bar{\nu}_l \rightarrow \bar{\nu}_{l'}$
oscillations in vacuum\cite{Cabibbo:1977nk,Bilenky:1980cx},
$P(\nu_l \rightarrow \nu_{l'})$ and
$P(\bar{\nu}_l \rightarrow \bar{\nu}_{l'})$,
$l\neq l' =e,\mu,\tau$,
can be caused, as is well known, by the
Dirac phase present in the Pontecorvo,
Maki, Nakagawa and Sakata (PMNS)
neutrino mixing matrix $U$.
If the neutrinos with definite masses
$\nu_i$, $i=1,2,3$, are Majorana particles,
the 3-neutrino mixing matrix contains
two additional Majorana CPV phases\cite{Bilenky:1980cx}. 
However, the flavour neutrino
oscillation probabilities
$P(\nu_l \rightarrow \nu_{l'})$ and
$P(\bar{\nu}_l \rightarrow \bar{\nu}_{l'})$,
$l,l' =e,\mu,\tau$, do not depend on
the Majorana phases\cite{Bilenky:1980cx,Langacker:1986jv}.
Our interest in the CPV
phases present in the neutrino mixing matrix
is stimulated also by the intriguing possibility
that the Dirac phase and/or the Majorana phases in
$U$ can provide the CP violation
necessary for the generation of the observed
baryon asymmetry of the Universe
\cite{Pascoli:2006ie,Pascoli:2006ci} 
(see also, e.g., Refs.~\refcite{Davidson:2008bu,Branco:2011zb}).

In the framework of the reference 3-flavour neutrino mixing
we will consider,
the PMNS neutrino mixing matrix
is always given by
$U = U_e^{\dagger} U_{\nu}$,
where $U_e$ and $U_{\nu}$ are $3 \times 3$ unitary matrices
originating from the diagonalisation of the charged lepton and
the neutrino (Majorana) mass terms.
We will suppose in what follows that
$U_{\nu}$ has a form which is dictated
by, or associated with, symmetries 
(see, e.g., Refs.~\refcite{Altarelli:2010gt,King:2013eh}).
In the present article we
consider the following symmetry forms of $U_{\nu}$:
i) tri-bimaximal (TBM)\cite{Harrison:2002k,Xing:2002sw,Wolfenstein:1978uw},
ii) bimaximal (BM)
(or corresponding to the conservation of the
lepton charge $L' = L_e - L_\mu - L_{\tau}$ (LC))\cite{Petcov:1982ya,
Vissani:1997pa,Barger:1998ta,Baltz:1998ey},
iii) golden ratio type A (GRA)\cite{Everett:2008et,Kajiyama:2007gx},
iv) golden ratio type B (GRB)\cite{Rodejohann:2008ir},
and v) hexagonal (HG)\cite{Albright:2010ap}.
For all these symmetry forms $U_\nu$ can be written as
\begin{equation}
U_\nu  = \Psi_1\,\tilde{U}_{\nu}\,Q_0 =
\Psi_1\,R_{23}(\theta^\nu_{23})
R_{12}(\theta^\nu_{12}) Q_0\,,
\label{UnuQ0}
\end{equation}
%
where $R_{23}(\theta^\nu_{23})$ and $R_{12}(\theta^\nu_{12})$ are
orthogonal matrices describing rotations in the 2-3 and 1-2 planes,
respectively, and $\Psi_1$ and $Q_0$ are diagonal phase matrices
each containing two phases. The phases in the matrix $Q_0$
give contribution to the Majorana phases in the PMNS matrix.
The symmetry forms of $\tilde{U}_{\nu}$ of interest,
TBM, BM (LC), GRA, GRB and HG, are
characterised by the same values of the
angles  $\theta^\nu_{13} = 0$ and
 $\theta^\nu_{23} = -\pi/4$,
but correspond to different fixed values of
the angle $\theta^\nu_{12}$ and
thus of $\sin^2\theta^\nu_{12}$, namely, to
i)  $\sin^2\theta^{\nu}_{12} = 1/3$ (TBM),
ii)  $\sin^2\theta^{\nu}_{12} = 1/2$ (BM (LC)),
iii)  $\sin^2\theta^{\nu}_{12} =  (2 + r)^{-1} \cong 0.276$ (GRA),
$r$ being the golden ratio, $r = (1 +\sqrt{5})/2$,
iv) $\sin^2\theta^{\nu}_{12} = (3 - r)/4 \cong 0.345$ (GRB), and
v) $\sin^2\theta^{\nu}_{12} = 1/4$ (HG).
The best fit values (b.f.v.) and $1\sigma$ errors 
of the three corresponding neutrino mixing parameters
in the standard parametrisation of the PMNS matrix\cite{Agashe:2014kda},
which we will employ,
read\cite{Capozzi:2013csa}: 
\begin{align}
\sin^2\theta_{12} &= 0.308^{+0.017}_{-0.017}\,,\\[0.30cm] 
\sin^2\theta_{13} &= 0.0234^{+0.0020}_{-0.0019}\,,\\[0.30cm]
\sin^2\theta_{23} &= 0.437^{+0.033}_{-0.023}\,, 
\end{align}
%
where the quoted values correspond to neutrino mass spectrum with 
normal ordering (NO); the values for spectrum 
with inverted ordering (IO) 
found in\cite{Capozzi:2013csa} differ insignificantly.
The minimal form of $U_e$ of interest
that can provide the requisite corrections to $U_{\nu}$,
so that the neutrino mixing angles
$\theta_{13}$, $\theta_{23}$ and  $\theta_{12}$
have values compatible with the current data,
including a possible sizeable deviation of $\theta_{23}$ from $\pi/4$,
includes a product of two orthogonal matrices describing
rotations in the 2-3 and 1-2 planes \cite{Marzocca:2013cr},
$R_{23}(\theta^e_{23})$ and $R_{12}(\theta^e_{12})$,
$\theta^e_{23}$ and $\theta^e_{12}$ being two (real) angles 
\footnote{For a detailed discussion of alternative possibilities see 
Ref. \refcite{Girardi:2015vha}.}.
This leads to the following parametrisation of the PMNS matrix $U$:
\begin{align}
U & =
R_{12} \left( \theta^e_{12} \right)\,
R_{23} \left( \theta^e_{23} \right)\,
\Psi\, R_{23} \left( \theta^\nu_{23} \right)
R_{12} \left( \theta^\nu_{12} \right) Q_0 \,,
\label{eq:U}
\end{align}
%
where $\Psi = {\rm diag} \left(1,e^{-i\psi}, e^{-i\omega} \right)$,
and $\theta^\nu_{23} = -\,\pi/4$. 
Equation (\ref{eq:U}) can be recast in the form \cite{Marzocca:2013cr}:
\begin{align}
U & =R_{12}(\theta^e_{12})\Phi(\phi)R_{23}(\hat\theta_{23})\,
R_{12}(\theta^{\nu}_{12})\,\hat{Q}\,,
\end{align}
%
where we have defined 
$\Phi = {\rm diag} \left(1,\text{e}^{i\phi},1\right)$, $\phi$
being a CPV phase, $\hat\theta_{23}$
is a function of $\theta^e_{23}$,
$\sin^2\hat\theta_{23} =
1/2 - \sin\theta^e_{23}\cos \theta^e_{23}\cos(\omega -\psi)$,
and $\hat{Q}$ is a diagonal phase matrix.
The phases in  $\hat{Q}$ give contributions
to the Majorana phases in the PMNS matrix.
The angle $\hat\theta_{23}$, however, can be expressed
in terms of the angles $\theta_{23}$ and $\theta_{13}$
of the PMNS matrix:
\begin{align}
\sin^2 \theta_{23} &=
\frac{\left| U_{\mu3} \right|^2}{1- \left| U_{e 3} \right|^2 }
= \frac{\sin^2\hat\theta_{23} - \sin^2 \theta_{13}}
{1 - \sin^2\theta_{13}}\,, 
\end{align}
%
and the value
of  $\hat\theta_{23}$ is fixed by the values of
$\theta_{23}$ and $\theta_{13}$.

\vspace{-0.3cm}
\bodymatter
%
\section{Predicting the Dirac Phase in the PMNS Matrix}
\label{aba:sec2}
%
%

 In the scheme under discussion the four observables
$\theta_{12}$, $\theta_{23}$, $\theta_{13}$
and  the Dirac phase $\delta$ in the PMNS matrix
are functions of three parameters
$\theta^e_{12}$, $\hat\theta_{23}$ and $\phi$.
As a consequence, the Dirac phase $\delta$
can be expressed as a function of the
three PMNS angles $\theta_{12}$, $\theta_{23}$
and $\theta_{13}$, leading to a new ``sum rule''
relating $\delta$ and $\theta_{12}$, $\theta_{23}$
and $\theta_{13}$. Within the approach employed 
this sum rule is exact.
Its explicit form depends on the symmetry
form of the matrix $\tilde{U}_{\nu}$, i.e.,
on the value of the angle $\theta^{\nu}_{12}$.
For arbitrary fixed value of $\theta^{\nu}_{12}$
the sum rule of interest reads \cite{Petcov:2014laa}: 
\begin{align}
\cos\delta =  \frac{\tan\theta_{23}}{\sin2\theta_{12}\sin\theta_{13}}\,
\left [\cos2\theta^{\nu}_{12} + 
\left (\sin^2\theta_{12} - \cos^2\theta^{\nu}_{12} \right )\,
 \left (1 - \cot^2\theta_{23}\,\sin^2\theta_{13}\right )\right ]\,.
\label{sumrule}
\end{align}
%
A similar  sum rule can be derived for 
the phase $\phi$~~\cite{Petcov:2014laa}.

 In Refs.~\refcite{Petcov:2014laa,Girardi:2014faa}
 we have derived predictions for 
$\cos\delta$, $\delta$ and the rephasing invariant 
$J_{\rm CP}$, which controls the magnitude of the CPV
effects in neutrino oscillations\cite{Krastev:1988yu}, 
using the sum rule in eq. (\ref{sumrule}) and the measured values of 
the lepton mixing angles $\theta_{12}$, $\theta_{13}$ and $\theta_{23}$.
In the present article we first summarise 
the predictions for these observables 
obtained in Refs.~\refcite{Petcov:2014laa,Girardi:2014faa}
in a simplified analysis employing the best fit values 
(b.f.v.)  and the $3\sigma$ allowed ranges of 
the three relevant neutrino mixing parameters 
$\sin^2\theta_{12}$, $\sin^2\theta_{13}$ and $\sin^2\theta_{23}$.  
This is followed by a summary of the results of the 
statistical analysis of the predictions performed 
in Ref.~\refcite{Girardi:2014faa}, which is 
based on i) the current, and most importantly, 
ii) the prospective, 
uncertainties in the measured values of $\sin^2\theta_{12}$, $\sin^2\theta_{13}$ 
and $\sin^2\theta_{23}$.

 We note first that the predicted values of $\cos\delta$
vary significantly with the symmetry form of 
$\tilde{U}_{\nu}$~\cite{Petcov:2014laa}.
For the best fit values of
  $\sin^2\theta_{12}=0.308$, $\sin^2\theta_{13}= 0.0234$ and
$\sin^2\theta_{23}= 0.437$~found in\cite{Capozzi:2013csa}, for instance,
we get $\cos\delta = (-0.0906)$, $(-1.16$),
$0.275$, $(-0.169)$ and $0.445$ for
the TBM, BM (LC), GRA, GRB and HG forms, respectively.
For the TBM, GRA, GRB and HG forms
these values correspond to  
$\delta = \pm 95.2^{\circ}, \pm 74.0^{\circ}, \pm 99.7^{\circ}, \pm 63.6^{\circ}$,
respectively.
The unphysical value of $\cos\delta$
in the  BM (LC) case  is a reflection of the fact that
the scheme under discussion with BM (LC)
form of the matrix $\tilde{U}_{\nu}$
does not provide a good description of the current data on
$\theta_{12}$, $\theta_{23}$ and $\theta_{13}$~
\cite{Marzocca:2013cr}.
Physical values of $\cos \delta$ can be obtained, 
for instance, for the b.f.v. of $\sin^2 \theta_{13}$
and $\sin^2 \theta_{23}$ if $\sin^2 \theta_{12}$
has a larger value\cite{Girardi:2014faa}: 
for, e.g.,  $\sin^2 \theta_{12} = 0.34$
allowed at $2\sigma$ by the current data,
we have $\cos \delta = -0.943$, 
corresponding to $\delta = \pm 160.6^{\circ}$.
Similarly, for  $\sin^2 \theta_{12} = 0.32$,  $\sin^2 \theta_{23}=0.41$ and 
$\sin\theta_{13}=0.158$ we have\cite{Petcov:2014laa}:
 $\cos \delta = -0.978$, $\delta = \pm 168.1^{\circ}$.

The results quoted above imply~\cite{Petcov:2014laa}
that the measurement of $\cos\delta$ can allow  to
distinguish between the different symmetry forms of
$\tilde{U}_{\nu}$, provided 
$\theta_{12}$, $\theta_{13}$ and $\theta_{23}$ 
are known with a
sufficiently good precision.
Even determining the sign of $\cos\delta$
will be sufficient to eliminate some of
the possible symmetry forms of $\tilde{U}_{\nu}$.

 It was also found  in\cite{Girardi:2014faa} that the sum 
rule predictions for
$\cos\delta$ exhibit strong dependence on the value of
$\sin^2\theta_{12}$ when the latter is varied in
its $3\sigma$ experimentally allowed 
range\cite{Capozzi:2013csa} (0.259\,--\,0.359).
The predictions for $\cos\delta$ change significantly
not only in magnitude, but also the sign of
$\cos\delta$ changes in the TBM, GRA and GRB cases 
\cite{Girardi:2014faa}.
In the case of $\theta^e_{23} = 0$, for instance,
we get for the TBM form of $\tilde{U}_{\nu}$
for the three values of
 $\sin^2 \theta_{12} = 0.308$, $0.259$ and $0.359$:
$\cos \delta = (-0.114)$, $(-0.469)$ and $0.221$,
thus $\cos\delta =0$ is allowed for a certain
value of  $\sin^2 \theta_{12}$.
 For the GRA and GRB forms of $\tilde{U}_{\nu}$
we have, respectively,
$\cos \delta = 0.289$, $(-0.044)$, $0.609$, and
$\cos \delta = (-0.200)$, $-0.559$, $0.138$.
Similarly, for the HG form we find
for the three values of $\sin^2 \theta_{12}$:
$\cos \delta = 0.476$, 0.153, 0.789.

In what concerns the dependence of 
the sum rule predictions for $\cos\delta$
when $\sin^2\theta_{23}$ is varied 
in its $3\sigma$ allowed interval,
$0.374\leq \sin^2\theta_{23}\leq 0.626$,
the results we obtained
 for $\sin^2\theta_{23} =0.374$ and
$\sin^2\theta_{23} =0.437$, setting
$\sin^2\theta_{12}$ and $\sin^2\theta_{13}$
to their best fit values, do not differ significantly.
However, the differences between the predictions
for $\cos\delta$ obtained
for  $\sin^2 \theta_{23} = 0.437$ and
for $\sin^2 \theta_{23} = 0.626$ are rather large\cite{Girardi:2014faa}
~---~they
differ by the factors of 2.05, 1.25, 1.77 and 1.32
in the TBM, GRA, GRB and HG cases, respectively.

 Similar analysis can be performed for the predictions for the
cosine of the phase $\phi$~~\cite{Girardi:2014faa}
which in many theoretical models 
serves as a ``source'' for the Dirac phase $\delta$.
 The phase $\phi$ is related to,
 but does not coincide with, the Dirac phase $\delta$.
 This leads to the confusing
 identification of $\phi$  with $\delta$:
 the sum rules satisfied by $\cos\phi$ and
 $\cos\delta$ differ significantly\cite{Petcov:2014laa}. 
Correspondingly,
 the predicted values of $\cos\phi$ and $\cos\delta$
 in the cases of the  TBM, GRA, GRB and HG
 symmetry forms of $\tilde{U}_{\nu}$ considered by us
 also differ significantly.
 This conclusion is not valid for the
 BM (LC) form: for this form the
 sum rules predictions for
 $\cos\phi$ and $\cos\delta$ are
 rather similar\cite{Petcov:2014laa}. 

We next present results of the statistical analysis
of the predictions for $\delta$, $\cos\delta$ and the rephasing
invariant $J_{\rm CP}$ performed in Ref.~\refcite{Girardi:2014faa}
in the cases of the TBM, BM (LC), GRA, GRB and HG symmetry
forms of the matrix $\tilde{U}_{\nu}$.
In this analysis 
the latest results on $\sin^2\theta_{12}$, $\sin^2\theta_{13}$,
$\sin^2\theta_{23}$ and $\delta$, obtained in the global analysis of the
neutrino oscillation data performed in \cite{Capozzi:2013csa} 
were used as input.
The aim was to derive the allowed ranges
for $\cos\delta$ and $J_{\rm CP}$,
predicted on the basis of the current data on
the neutrino mixing parameters for each of the
symmetry forms of $\tilde{U}_{\nu}$ considered.
For this purpose 
the $\chi^2$ function was constructed 
in the following way\cite{Girardi:2014faa}:
$\chi^2(\{x_i\}) = \sum_i \chi_i^2(x_i)$,
with $x_i = \{\sin^2 \theta_{12},\sin^2 \theta_{13},\sin^2 \theta_{23},\delta\}$.
The functions $\chi^2_i$ have been extracted from 
the 1-dimensional projections given in Ref.~\refcite{Capozzi:2013csa} and, thus 
the correlations between the oscillation parameters have been neglected.
This approximation is sufficiently precise since it allows 
to reproduce the contours in the planes 
$(\sin^2 \theta_{23} , \delta)$, $(\sin^2 \theta_{13} , \delta)$ 
and $(\sin^2 \theta_{23} , \sin^2\theta_{13})$, given in\cite{Capozzi:2013csa}, 
with a rather high accuracy. We 
calculated $\chi^2 (\cos \delta)$ by marginalising 
$\chi^2(\{x_i\})$ over $\sin^2 \theta_{13}$ and $\sin^2 \theta_{23}$ 
for a fixed value of $\cos \delta$.
Given the global fit results, the likelihood function,
\begin{align}
L(\cos \delta) \propto \exp \left(- \frac{\chi^2 (\cos \delta)}{2} \right) \,,
\end{align}
%
represents the most probable value of $\cos \delta$ for
each of the considered symmetry forms of $\tilde U_{\nu}$.
The $n \sigma$ confidence level (C.L.) region corresponds to the
interval of values of $\cos \delta$ in which
$L(\cos \delta) \geq L(\chi^2 = \chi^2_{\rm min}) \cdot L(\chi^2 = n^2)$,
where $\chi^2_{\rm min}$ is the value of $\chi^2$
in the minimum.

\begin{figure}[|t]
\centering
\includegraphics[width=\columnwidth]{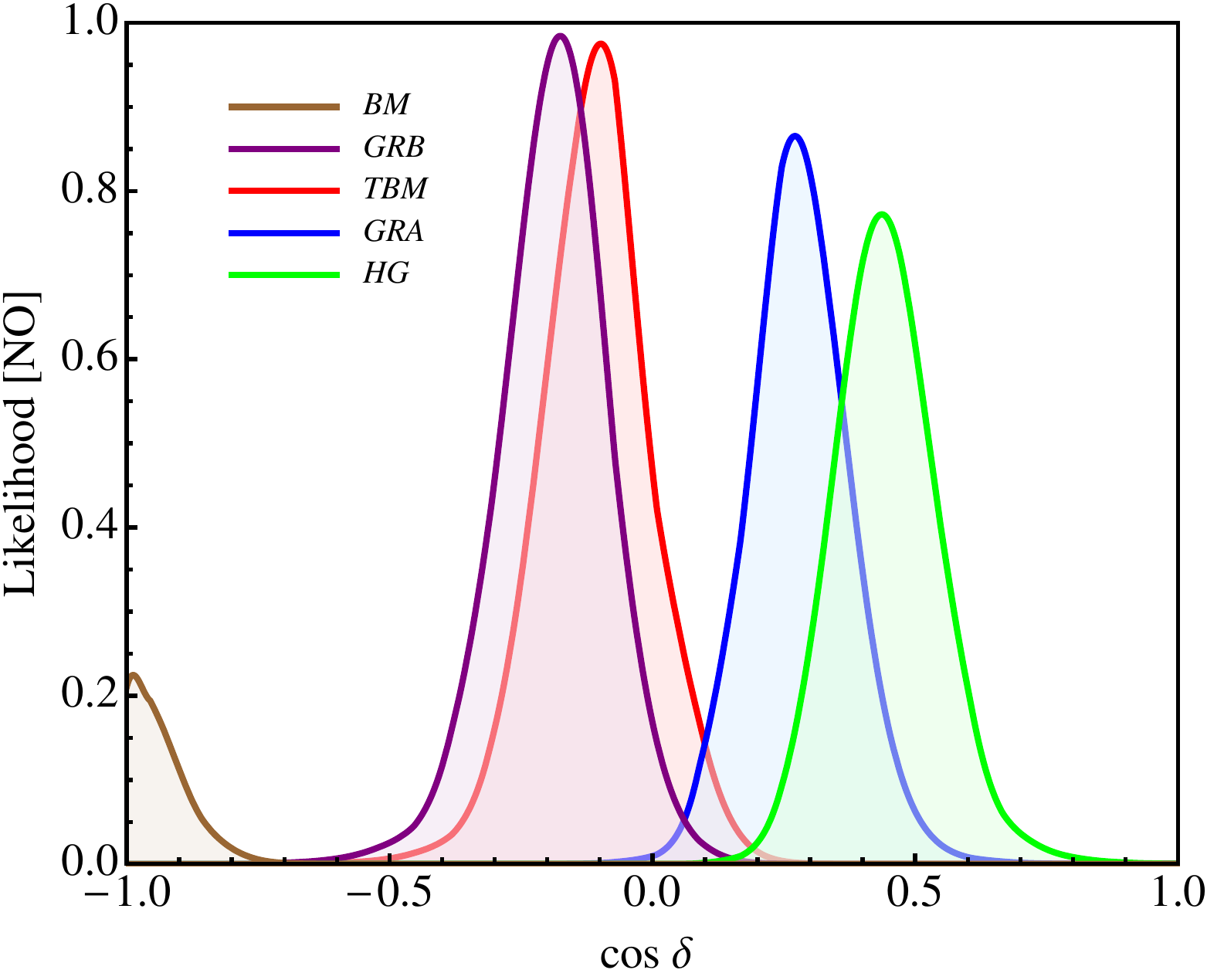}
\vspace{-0.6cm}
\caption{The likelihood function
versus $\cos\delta$ for NO
neutrino mass spectrum after marginalising over
$\sin^2\theta_{13}$ and $\sin^2\theta_{23}$,
for the TBM, BM (LC), GRA, GRB and HG symmetry forms
of the mixing matrix $\tilde{U}_{\nu}$
(see text for further details). (From 
Ref.~\protect\refcite{Girardi:2014faa}.) 
}
\label{Fig:1}
\end{figure}

In Fig.~\ref{Fig:1} we show the
likelihood function versus $\cos \delta$ for NO
neutrino mass spectrum from Ref.~\refcite{Girardi:2014faa}. 
The results shown are obtained
by marginalising over all the
other relevant parameters of the scheme considered.
The dependence of the likelihood function on $\cos \delta$
in the case of IO neutrino mass spectrum differs little 
from that shown in Fig.~\ref{Fig:1}.
%
As can be observed in Fig.~\ref{Fig:1},
a rather precise measurement of $\cos \delta$ would allow 
to distinguish between the different symmetry forms of $\tilde U_{\nu}$
considered by us. For the TBM and GRB forms
there is a significant overlap of the corresponding likelihood functions.
The same observation is valid for the GRA and HG forms.
However, the overlap of the likelihood functions of these
two groups of symmetry forms occurs only at $3\sigma$ level
in a very small interval of values of $\cos \delta$.
This implies that in order to distinguish
between TBM/GRB, GRA/HG and BM (LC) symmetry forms, a
not very demanding measurement (in terms of accuracy) of $\cos \delta$
might be sufficient. The value of the non-normalised likelihood
function at the maximum in Fig.~\ref{Fig:1}
is equal to $\exp( - \chi^2_{\rm min}/2)$, which
allows us to make conclusions
about the compatibility of the symmetry schemes
considered with the current global data.
The results of this analysis for $\cos \delta$
are summarised in Table~\ref{tab:Fig1}.

We have also performed in Ref.~\refcite{Girardi:2014faa} 
a similar statistical analysis
of the predictions for the rephasing
invariant $J_{\rm CP}$  
in the cases of the TBM, BM (LC), GRA, GRB and HG symmetry
forms of the matrix $\tilde{U}_{\nu}$ considered.
In this analysis we used as input
the latest results on $\sin^2\theta_{12}$, $\sin^2\theta_{13}$,
$\sin^2\theta_{23}$ and $\delta$, obtained in the global analysis of the
neutrino oscillation data performed in Ref.~\refcite{Capozzi:2013csa},
and minimised $\chi^2$ for a fixed value of $J_{\rm CP}$.
The obtained b.f.v. and $3\sigma$ ranges
are given in Table \ref{tab:Fig1}.
We have found, in particular, that
the CP-conserving value of $J_{\rm CP} = 0$ is
excluded  in the cases of the  TBM, GRA, GRB
and HG neutrino mixing symmetry forms, respectively,
at approximately $5\sigma$, $4\sigma$, $4\sigma$
and $3\sigma$ C.L. with respect to the C.L.
of the corresponding best fit value.
 These results reflect the predictions
we have obtained for $\delta$,
more specifically, the C.L. at which
the CP-conserving values of  $\delta = 0~(2\pi)$, $\pi$,
are excluded in the discussed cases.
We found also that the $3\sigma$
allowed intervals of values of
$\delta$ and $J_{\rm CP}$ are rather narrow for
all the symmetry forms considered,
except for the BM (LC) form.
More specifically, for the TBM, GRA, GRB
and HG symmetry forms we have obtained
at $3\sigma$: $0.020 \leq |J_{\rm CP}| \leq 0.039$.
For the b.f.v. of $J_{\rm CP}$ we have found,
respectively:
$J_{\rm CP} = (-0.034)$, $(-0.033)$, $(-0.034)$, and  $(-0.031)$.
Our results indicate that distinguishing between
the TBM, GRA, GRB and HG symmetry forms of the neutrino mixing
would require extremely high precision measurement of
the $J_{\rm CP}$ factor~\cite{Petcov:2014laa}.

\begin{figure}[!t]
\centering
\includegraphics[width=\columnwidth]{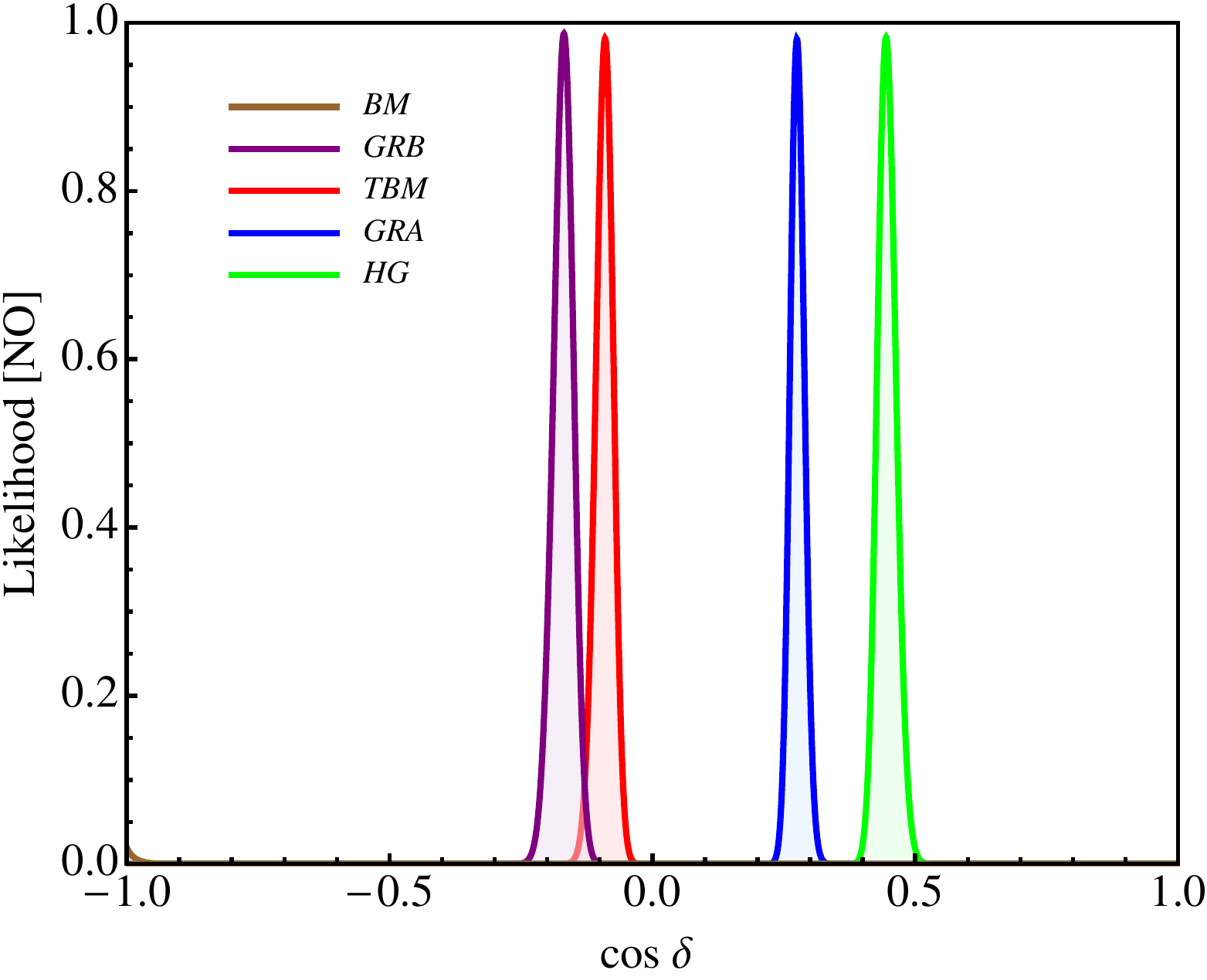}
\vspace{-0.6cm}
\caption{The same as in Fig.~\ref{Fig:1},
but using the prospective $1\sigma$ uncertainties in the 
determination of $\sin^2\theta_{12}$, $\sin^2\theta_{13}$ and 
$\sin^2\theta_{23}$
within the Gaussian approximation.
The three neutrino mixing parameters 
are fixed to their current best fit values
(i.e., $\sin^2 \theta_{12} = 0.308$, etc.). 
See text for further details.
(From Ref.~\protect\refcite{Girardi:2014faa}.)
}
\label{Fig:2}
\end{figure}

 In Fig.~\ref{Fig:2} we present the likelihood function
versus $\cos \delta$ within the Gaussian approximation,
i.e., using $\chi^2_{\rm G} = \sum_i  (y_i - \overline y_i)^2 / \sigma^2_{y_i}$, 
with $y_i = \{\sin^2 \theta_{12},\sin^2 \theta_{13},\sin^2 \theta_{23}\}$, 
where we used the current b.f.v. ($\overline y_i$)
of the mixing angles for NO neutrino mass spectrum given in 
Ref.~\refcite{Capozzi:2013csa} and the prospective 
$1\sigma$ uncertainties ($\sigma_{y_i}$) in the determination of 
$\sin^2 \theta_{12}$ (0.7\% from JUNO \cite{Wang:2014iod}),
$\sin^2 \theta_{13}$ (3\% derived from an expected error on 
$\sin^2 2 \theta_{13}$ of 3\% from Daya Bay, 
see Refs.~\refcite{deGouvea:2013onf,Zhang:2015fya}) 
and $\sin^2 \theta_{23}$ (5\% derived from the potential sensitivity of 
NOvA and T2K on $\sin^2 2 \theta_{23}$ 
\begin{table}[t]
\tbl{Best fit values (b.f.v.) of $J_{\rm CP}$
and $\cos \delta$
and corresponding 3$\sigma$ ranges
(found fixing $\chi^2-\chi^2_{\rm min} = 9$)
in our setup
using the data from\cite{Capozzi:2013csa}
for NO neutrino mass spectrum.
(From Ref. \protect\refcite{Girardi:2014faa}, where results for 
IO spectrum are also given.)
}
{
\begin{tabular}{lcccc}
\toprule
Scheme & $J_{\rm CP}/10^{-2}$ (b.f.v.) & $J_{\rm CP}/10^{-2}$ ($3\sigma$ range) & $\cos \delta$ (b.f.v.) & $\cos \delta$ ($3\sigma$ range)\\
\colrule
TBM & $-3.4$ & $[-3.8,-2.8] \cup [3.1,3.6]$ & $-0.07$ & $[-0.47,\phantom{-}0.21]$\\
BM (LC) & $-0.5$ & $[-2.6,2.1]$ & $-0.99$ & $[-1.00,-0.72]$\\
GRA & $-3.3$ & $[-3.7,-2.7] \cup [3.0,3.5]$ & $\phantom{-}0.25$ & $[-0.08,\phantom{-}0.69]$\\
GRB & $-3.4$ & $[-3.9,-2.6] \cup [3.1,3.6]$ & $-0.15$ & $[-0.57,\phantom{-}0.13]$\\
HG & $-3.1$ & $[-3.5,-2.0] \cup [2.6,3.4]$ & $\phantom{-}0.47$ & $[\phantom{-}0.16,\phantom{-}0.80]$\\
\botrule
\end{tabular}}
\label{tab:Fig1}
\end{table} 
%
\noindent of 2\%, see Ref.~\refcite{deGouvea:2013onf}, this sensitivity 
can be also achieved in future neutrino facilities
as T2HK\cite{Coloma:2014kca}).
The BM (LC) case is very sensitive to the b.f.v. of $\sin^2 \theta_{12}$
and $\sin^2 \theta_{23}$ and is disfavoured at more than $2\sigma$ for
the current b.f.v. found in \cite{Capozzi:2013csa}. 
This case might turn out to be compatible with the data 
for larger (smaller) measured values of 
$\sin^2 \theta_{12}$ ($\sin^2 \theta_{23}$), 
as can be seen from Fig.~\ref{Fig:3}, which was obtained
for $\sin^2 \theta_{12} = 0.332$
(the best fit values of the two other mixing angles being 
kept intact). With the increase of the value of $\sin^2 \theta_{23}$
the BM (LC) form becomes increasingly disfavoured, 
while the TBM/GRB (GRA/HG) predictions
for $\cos \delta$ are shifted somewhat
to the left (right) with respect 
to those shown in Fig.~\ref{Fig:2}. 
For, e.g.,  the best fit values of  
$\sin^2 \theta_{12} = 0.304$, 
$\sin^2 \theta_{13} = 0.0219$ and 
$\sin^2 \theta_{23} = 0.579$,  
found in Ref.~\refcite{Gonzalez-Garcia:2014bfa} 
for IO neutrino mass spectrum, 
these shifts in $\cos\delta$ are approximately by 0.1.

The measurement of $\sin^2 \theta_{12}$, $\sin^2 \theta_{13}$
and $\sin^2 \theta_{23}$ with the quoted precision will open up
the possibility to distinguish between the BM (LC), TBM/GRB, GRA and HG forms of
$\tilde U_{\nu}$. Distinguishing between the TBM and GRB forms would
require relatively high precision measurement of $\cos \delta$.
Assuming that $|\cos \delta| < 0.93$, which means for 
76\% of values of $\delta$, the error on $\delta$, $\Delta \delta$,
for an error on $\cos \delta$, $\Delta(\cos \delta) = 0.10 \, (0.08)$,
does not exceed $\Delta \delta \lesssim 
\Delta(\cos \delta) /\sqrt{1 - 0.93^2} = 16^{\circ} \, (12^{\circ})$.
This accuracy is planned to be reached in the future neutrino experiments
like T2HK (ESS$\nu$SB) \cite{deGouvea:2013onf}.
Therefore a measurement of $\cos \delta$ in the quoted range 
will allow one to distinguish between the TBM/GRB, BM (LC) and
GRA/HG forms at approximately $3\sigma$ C.L.,
if the precision achieved on $\sin^2 \theta_{12}$,
$\sin^2 \theta_{13}$ and $\sin^2 \theta_{23}$ is the same
as in Figs.~\ref{Fig:2} and \ref{Fig:3}.
%

\vspace{-0.1cm}
\bodymatter
%
\section{Summary and Conclusions}
\label{aba:sec2}
%
%
In conclusions, we have derived in\cite{Girardi:2014faa} 
the ranges of the predicted values of 
$\cos \delta$ and $J_{\rm CP}$ for the TBM, BM (LC), GRA, GRB and HG 
symmetry forms of $\tilde U_{\nu}$, 
from a statistical analysis using the 
sum rule in eq. (\ref{sumrule}) obtained  
in\cite{Petcov:2014laa} and the
current global neutrino oscillation data \cite{Capozzi:2013csa}.
The results of this analysis are summarised in Table \ref{tab:Fig1}
and in Fig.~\ref{Fig:1}.
We found, in particular, that in the 
TBM, GRA, GRB and HG cases, the 
best fit values of $J_{\rm CP}$ lie in the narrow interval
$(-0.034) \leq J_{\rm CP}\leq (-0.031)$, while
at $3\sigma$ we have $0.020 \leq |J_{\rm CP}| \leq 0.039$.
The predictions for $\delta$, $\cos\delta$ and $J_{\rm CP}$
in the case of the BM (LC) symmetry form of $\tilde{U}_{\nu}$,
as the results of the statistical analysis performed by us showed,
differ significantly:
the best fit value of $\delta \cong \pi$, and, correspondingly,
of $J_{\rm CP} \cong 0$. For the $3\sigma$ range 
in the case of NO (IO) neutrino mass spectrum we find:
$-0.026~(-0.025)\leq J_{\rm CP} \leq 0.021~(0.023)$, i.e.,
it includes a sub-interval of values centred on zero, which does not
overlap with the $3\sigma$ allowed intervals of values
of $J_{\rm CP}$, corresponding to the TBM, GRA, GRB and HG 
symmetry forms of  $\tilde U_{\nu}$.

\begin{figure}[!t]
\centering
\includegraphics[width=\columnwidth]{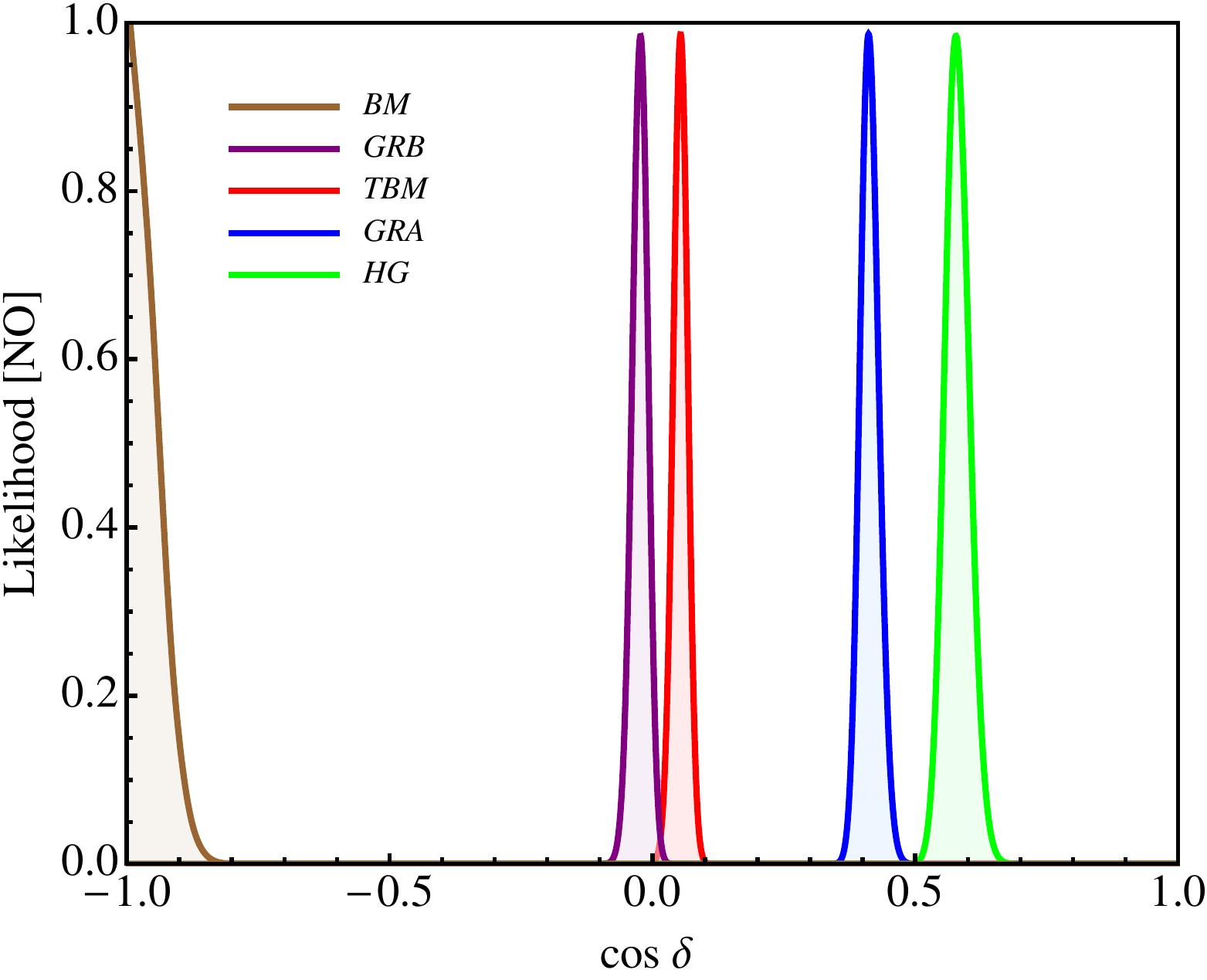}
\vspace{-0.6cm}
\caption{The same as in Fig.~\ref{Fig:2},
but using $\sin^2 \theta_{12} = 0.332$. 
(From Ref.~\protect\refcite{Girardi:2014faa}.)
}
\label{Fig:3}
\end{figure}

 Finally, we have derived in\cite{Girardi:2014faa} 
predictions for $\cos \delta$
using the prospective 
$1\sigma$ uncertainties in the determination of
$\sin^2 \theta_{12}$, $\sin^2 \theta_{13}$ and $\sin^2 \theta_{23}$
respectively in JUNO, Daya Bay and accelerator and atmospheric
neutrino experiments (Figs.~\ref{Fig:2} and \ref{Fig:3}).
The results thus obtained show that i) the measurement of 
the sign of $\cos \delta$ will allow to distinguish between the TBM/GRB, BM (LC)
and GRA/HG forms of $\tilde U_{\nu}$, ii) for a best fit value
of $\cos \delta = -1 \, (-0.1)$ distinguishing at $3\sigma$ between 
the BM (TBM/GRB) and the other forms of $\tilde U_{\nu}$
would be possible if $\cos \delta$ is measured with $1\sigma$
uncertainty of $0.3 \, (0.1)$. 

The results obtained in the studies performed in 
Refs.~\refcite{Petcov:2014laa,Girardi:2014faa}
show, in particular, that the experimental measurement of 
the Dirac phase $\delta$ of the PMNS neutrino mixing matrix
in the future neutrino experiments,
combined with the data on the neutrino mixing angles
can provide unique information about the possible
discrete symmetry origin of the observed pattern of
neutrino mixing.

%

\vspace{0.2cm}
{\bf Acknowledgements.}
This work was supported in part by the European Union FP7
ITN INVISIBLES (Marie Curie Actions, PITN-GA-2011-289442-INVISIBLES),
by the INFN program on Theoretical Astroparticle Physics (TASP),
by the research grant  2012CPPYP7 ({\sl  Theoretical Astroparticle Physics})
under the program  PRIN 2012 funded by the Italian 
Ministry of Education, University and Research (MIUR)
and by the  World Premier International Research Center 
Initiative (WPI Initiative, MEXT), Japan (STP).

\vspace{-0.4cm}

\end{document}